\documentclass[a4paper]{mem}
\usepackage{natbib}
\usepackage{graphicx}
\usepackage[a4paper]{hyperref}
\idline{75}{1}
\begin{document}
   \title{Mid-infrared Photometry of Carbon Stars\\
    and Perspectives for Surveys in the\\
    Magellanic Clouds from Dome-C
}

   \author{Stefano Ciprini\inst{1,2} \and Maurizio Busso\inst{1,3}
}

   \offprints{S. Ciprini}
\mail{via Pascoli, 06123 Perugia}

   \institute{Dipartimento di Fisica e Osservatorio Astronomico, Universit\'a di
   Perugia,\\ via Pascoli, 06123 Perugia, Italy\\
                \email{stefano.ciprini@pg.infn.it~-~maurizio.busso@fisica.unipg.it}\\
              \and  INFN, Sezione di Perugia, via Pascoli, 06123 Perugia, Italy\\
              \and INAF Osservatorio Astronomico di Torino, 10025 Pino Torinese (TO), Italy
             }

   \abstract{A preliminary analysis of the data from the MSX space infrared satellite,
   seems to confirm that the [8.8]-[12.5] micron color index is well correlated with the mass-loss
   rates in carbon stars of our Galaxy. The extension of this mid-infrared observation criterion to
   the Magellanic Clouds, with a small-size telescope like IRAIT, able to perform a continuous survey
   from Dome-C on the Antarctic Plateau, could be crucial to trace the local AGB population
   and evolution.
   }
   \authorrunning{S. Ciprini \& M. Busso}
   \titlerunning{Mid-infrared Photometry of Carbon Stars}
   \maketitle
%

\section{Introduction}

The Midcourse Space
Experiment\footnote{\texttt{http://www.ipac.caltech.edu/ipac/msx/}}
\citep[MSX satellite;][]{mill94} was launched in April 1996, as an
USA ballistic missile defense project. The first ten months of the
mission were devoted to mid-infrared observations with a solid
hydrogen cooled telescope (SPIRIT III: Spatial Infrared Imaging
Telescope III). This instrument had focal plane infrared arrays
operated at 11 to 12 $^{\circ}$K by employing a solid hydrogen
cryostat, that spanned the spectral region from 4.2 to 26 microns,
with a beam-size 35 times smaller than IRAS, resulting in images
with excellent spatial resolution. The sensitivity in the MSX A
band (8.28 $\mu$m) was about four times then IRAS. The cryogen
phase of the mission ended on February 1997. A full set of
experiments mapped the entire Galactic Plane between +5$^{\circ}$
and -5$^{\circ}$ of latitude, the $\sim$4\% of the sky unobserved
by IRAS, the zodiacal background, confused regions away from the
Plane, deep surveys of selected fields at high galactic latitudes,
large galaxies, asteroids and comets. The data from Galactic Plane
and IRAS gaps surveys as well as observations of the LMC have been
processed by the Air Force Research Laboratory (AFRL) and an atlas
of images and a catalog of point sources have been produced. Over
200 Gb of data on Celestial Backgrounds were obtained during the
ten month cryogen phase of the mission (archive data at
IRSA\footnote{\texttt{http://irsa.ipac.caltech.edu/\\applications/MSX/}}).
   \begin{figure}[t!]
   \centering
   \resizebox{\hsize}{!}{\rotatebox[]{0}{\includegraphics{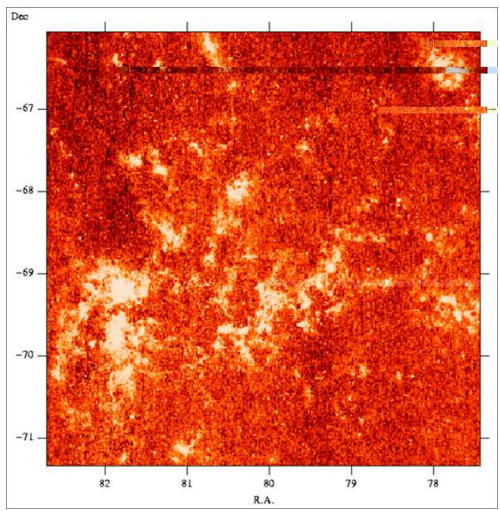}}}
   \caption{MSX Spirit III infrared mosaic image of the Large Magellanic Cloud in
   its broad A band (centered at 8.276$\mu$m with a bandwidth between 6.8$\mu$m and 10.8$\mu$m).}
              \label{fig:LMC}%
    \end{figure}
The MSX Infrared Point Source Catalog version 1.2 \citep{egan01b}
is accompanied by an Explanatory Guide describing the instruments,
data processing, calibration, the catalogs and the
photometric/positional accuracy \citep{egan99}. The validation
method for the absolute calibration of the SPIRIT III infrared
radiometry is described in \citep{cohen00}.
\par
The other data used in this work are taken from the last two
observing winter campaigns of the mid-infrared camera TIRCAM II
\citep[the upgraded version of the Tirgo mid-InfraRed CAMera, see
][]{persi01,persi94} at the Tirgo
Observatory\footnote{\texttt{http://www.arcetri.astro.it/irlab/\\tirgo/}}
\citep[Gornergrat Infrared Telescope, ][]{mannucci03}. The
instrument is endowed of a Rockwell high flux 128x128 Si:As
blocked-impurity-band (BIB) focal plane array, suited for the high
background condition typical of the ground based applications. In
perspective of the adaptation to IRAIT
\citep[Italian/International Robotic Antarctic Infrared
Telescope)][]{tosti03thispr,tosti03mem,busso02pasa}, the control
system has been revised both on hardware and software components
\citep{corcione03}.

   \begin{figure*}
   \centering
   \resizebox{10cm}{!}{\rotatebox[]{0}{\includegraphics{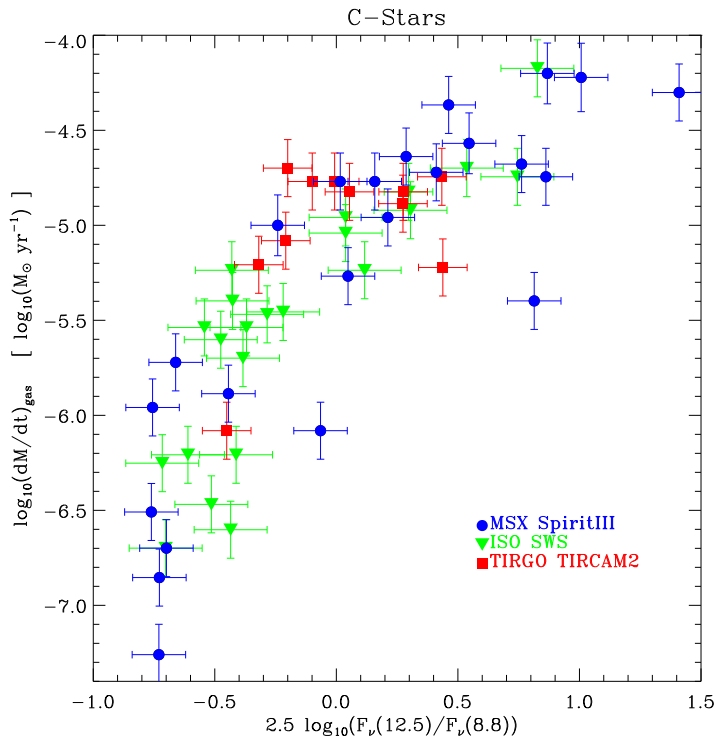}}}
   \caption{Color index to mass loss diagrams for our sample of C-rich stars.}
              \label{fig:massloss}%
    \end{figure*}

\section{The mid-IR color to mass-loss diagram}
A sample of carbon-rich stars in the Galactic Plane observed by
the MSX satellite is preliminary analyzed and showed in Fig.
\ref{fig:massloss}, in view of the development of photometric
criteria suitable to discriminate between O-rich and C-rich
sources. This criteria based on plain mid-infrared colors, should
let to derive mass-loss rates and predictions on the evolutionary
status of the observed sources, and the consequences for the mass
loss history on the Asymptotic Giant Branch (AGB) stars.
\par
In Fig. \ref{fig:massloss} we show a sample of carbon-rich stars.
A well defined relation holds between the mass loss of the gas and
the mid-infrared color index [8.8]-[12.5] ($\mu$m). In this
preliminary diagram we used the mid-infrared observations
performed during the last two winter observing campaigns of TIRCAM
II, and the data from the MSX database (simply taking the original
fluxes in its A and C bands). Mid-infrared ISO fluxes also
enclosed in Figure \label{fig:massloss} at this two bands, were
calculated by the good accuracy data of the Short Wavelength
Spectrometer (SWS) of this satellites, \citep{corti03a}.
\par
Diagram of Fig. \ref{fig:massloss} was constructed by using
mass-loss data compiled by
\citep{knapp98,fuente98,neri98,loup93,kastner93}. More distinct
references were used in the case of singles post-AGB and
pre-planetary  nebulae (PN) enclosed in the diagram. The analysis
with the new mass-loss data appeared in
\citet{groen02,olofsson02,lebertre01} and the mid-IR fluxes of MSX
on an extended sample of C-rich and S-rich AGB is an ongoing work.
\par
The carbon-rich stars in Fig. \ref{fig:massloss} are distributed
in a well defined branch (except for few cases of peculiar stars,
like Y Tau, or V Cyg, where mid-IR variability might have an
important role). Our preliminary diagram point out a relation,
appearing linear for intermediate values of mass-loss and color,
and which seems to continue toward higher wind velocities and
higher reddening, in the rather unexplored region of the post-AGB
and pre-PNs (top-right part of the plot). The stars in this
region, like for example V353 Aur (CRL 618, Westbrook Nebula), CW
Leo (Peanut Nebula), OHPN 10, RAFGL 2477, NGC 7027, V354 Lac, IRAS
21282-5050, indeed are all classified as pre-PNs or PNs and the
relation looks still valid.
\par
The power of the diagnostic tools based on mid-infrared colors was
underlined for example in \citep{busso96,marengo99}. The different
chemical signatures of carbon-rich and oxygen-rich envelopes are
put in evidence by mid-IR color-color diagrams, and correlations
between the observed colors and mass-loss rates appear clear
already in the first data obtained with TIRCAM in the previous
years. Moreover plain photometric mid-infrared observations do the
possibility to derive direct information on the spatial structure
and symmetry of the circumstellar envelopes.
   \begin{figure}[t!]
   \centering
   \resizebox{\hsize}{!}{\rotatebox[]{0}{\includegraphics{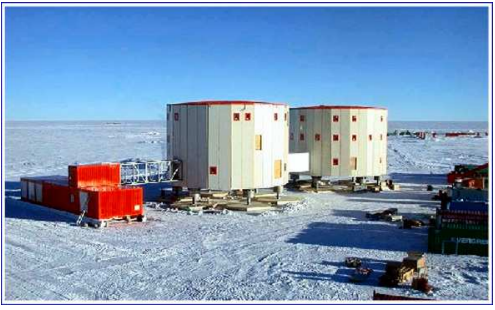}}}
   \caption{The two main buildings of the Concordia permanent station at Dome
   C (3280m a.s.l.) on the Antarctica Plateau, as appear at the end of the last summer
   expedition (Courtesy of M. Candidi).}
              \label{fig:domeC}%
    \end{figure}
\par
The possibly parabolic relation outlined by the plot, evidences
for possible temporal variations in the mass loss rates and an
increasing level of the [C/O] abundance going to the upper part of
the diagram branch. This might be represented by a semi-analytic
and time dependent function. Constraints come from the mass of the
final central white dwarfs, because if the circumstellar envelopes
are too compact, the mixing in the stellar atmosphere should be
not sufficient, obstructing the formation of low mass
carbon-stars. The understanding of the nucleosynthesis in low
luminosity carbon stars (enriched in s-elements) has recently
succeeded using high resolution codes with improved opacities,
which account for the third dredge-up. The models point out this
stars as the normal outcome of AGB evolution, characterized by
production of 12C and neutron-rich nuclei in the He intershell and
by mass loss from strong stellar winds \citep{busso99}. An
enlargement of the sample with new radio-millimeter estimations of
the mass-loss and MSX mid-infrared fluxes will be useful to probe
this relation and the diagnostic method.

\section{Dome C: perspectives for mid-IR surveys in Magellanic Clouds}
The French-Italian base of Dome-C \citep{candidi03}, sited on the
Antarctica Plateau at an altitude of 3280m asl will permit the
development of interesting astrophysical projects, because the
inner Plateau features the best sky conditions for ground-based
millimeter and infrared observations
\citep{calisse03,valenziano03,chamberlain00,hidas00,valenziano99}.
\par
Moderate size telescopes placed in this site, for example could
carry out a mid-IR photometric monitoring and survey of the
Magellanic Clouds. The 0.8m IRAIT telescope
\citep[Italian/International Robotic Antarctic Infrared Telescope,
][]{tosti03thispr,tosti03mem,busso02pasa}, in the first phase will
be operative at Dome C with the update version of TIRCAM II. An
analogous diagram might be constructed for the Magellanic Clouds
with simple mid-infrared photometric observations by IRAIT (for a
preliminary estimation of the IR flux limits and photometric
performances see \citet{fiorucci03}).
\par The relation between the mass-loss rates (derivable from millimeter
observations if the distance is known) and the [8.8]-[12.5] color
index, extended to the Magellanic Clouds of known distance
modulus, will allow to derive the directly $\dot{M}$ from a plain
mid-IR photometry. The Magellanic Clouds are dwarf irregular
galaxies which are in general metal-poor. Nebular abundances are
roughly one-third of the solar in the Large Magellanic Cloud
(LMC), and one-tenth of the solar in the Small Magellanic Cloud
(SMC). Mid-IR samples in this two galaxies from Antarctica, join
to the existing and extended Galactic samples (MSX), through the
accurate Hipparcos parallaxes, will permit to obtain three source
populations of different metallicity very important to study the
formations of the planetary nebulae and the return of material in
the interstellar medium. The pre-PN phase, between the C-stars of
[C/O]$\simeq$1 and the PNs, is a very unexplored field because
this stage is completely dominated by infrared emission and
invisible in the optical. Over 0.5${M}_{\sun}$ are believed loss
and about 50-70\% of the carbon is transferred to the Galaxy in
this phase by the star. This is a very important stage for AGBs
and galaxy evolution but almost unexplored, so the expectations
for the next mid-IR astronomy from Dome C cannot be anything else
than very good.

\bibliographystyle{aa}


\end{document}